\documentclass[a4paper, amsfonts, amssymb, amsmath, reprint, showkeys, footinbib, twoside,superscriptaddress,floatfix,longbibliography]{revtex4-1}

\usepackage{graphicx}%
\usepackage{dcolumn}%
\usepackage{bm}%
\usepackage{float}
\usepackage{xcolor}
\usepackage{mhchem}
\usepackage{gensymb}
\usepackage{ulem}
\usepackage{comment}
\usepackage[pdftex, bookmarks, pdffitwindow=false, pdfstartview=FitH, pdfdisplaydoctitle, colorlinks, plainpages=false, pdftitle={},pdfauthor={}, pdfpagelabels, hypertexnames, citecolor={blue!50!black},linkcolor={blue!50!black}, urlcolor={blue!50!black}, pdflang={en}, hyperfootnotes=false, breaklinks]{hyperref}
\newcommand{\si}{Supplementary Materials}
\newcommand{\avg}[1]{\ensuremath{\left<#1\right>}}

\DeclareMathOperator*{\argmin}{arg\,min}

\newcommand{\tm}{\ensuremath{T_{\text{m}}}} %
\newcommand{\kb}{k_{\rm B}} %

\makeatletter
\renewcommand{\@seccntformat}[1]{}
\makeatother

\begin{document}

\preprint{APS/123-QED}

\title{Predicting the phase behaviors of superionic water at planetary conditions}

\author{Bingqing Cheng}
\email{bc509@cam.ac.uk} 
\affiliation{Department of Computer Science and Technology, J.\,J.\,Thomson Avenue, Cambridge, CB3 0FD, United Kingdom}
\affiliation{Cavendish Laboratory, University of Cambridge, J.\,J.\,Thomson Avenue, Cambridge, CB3 0HE, United Kingdom}

\author{Mandy Bethkenhagen}
\affiliation{\'Ecole Normale Sup\'erieure de Lyon,  Universit\'e Lyon 1, Laboratoire de G\'eologie de Lyon, CNRS UMR 5276, 69364 Lyon Cedex 07, France}

\author{Chris J. Pickard}
\affiliation{Department of Materials Science \& Metallurgy, University of Cambridge, 27 Charles Babbage Road,
Cambridge, CB3 0FS, United Kingdom}
\affiliation{Advanced Institute for Materials Research, Tohoku University, Sendai, Japan}

\author{Sebastien Hamel}
\affiliation{Lawrence Livermore National Laboratory, Livermore, California 94550, USA}

\date{\today}%

\begin{abstract}
%
%
%
%
%
Most water in the universe may be superionic,
and its thermodynamic and transport properties are crucial for planetary science but difficult to probe experimentally or theoretically.
We use machine learning and free energy methods to overcome the limitations of quantum mechanical simulations, 
and characterize hydrogen diffusion, superionic transitions, and phase behaviors of water at extreme conditions.
We predict that a close-packed superionic phase with mixed stacking is stable over a wide temperature and pressure range,
while a body-centered cubic phase is only thermodynamically stable in a small window but is kinetically favored.
Our phase boundaries, which are consistent with the existing-albeit scarce-experimental observations, help resolve the fractions of insulating ice, different superionic phases, and liquid water inside of ice giants.
\end{abstract}

\maketitle

Water is the dominant constituent of Uranus'
and Neptune's mantle~\cite{Helled2020}, and superionic water is believed to be stable at depths greater than about one-third of the radius of these ice giants~\cite{Millot2019}.
Water superionicity is marked by exceptionally high hydrogen diffusivity and ionic conductivity,
as hydrogen atoms become liquid-like while oxygen atoms remain solid-like on a lattice with a certain crystallographic order. 
Although superionic water was postulated over three decades ago~\cite{demontis1988}, its optical properties (it is partially opaque) and oxygen lattices were only accurately measured recently~\cite{Millot2018,Millot2019}, %
and many properties of this hot ``black ice'' are still uncharted.

Amongst the many mysteries regarding superionic water,
the phase behaviors over a large range of pressure ($P$) and temperature ($T$) are particularly important in planetary science: The location of various coexistence lines 
(including the high-pressure melting line, the insulating ice to superionic transition line, 
and the phase boundaries between competing superionic phases of water) 
are essential for understanding the formation, evolution, interior structure and magnetic fields of planets~\cite{Redmer2011,Helled2020,Soderlund2020}
However, the locations of these coexistence lines and even the possible types of oxygen lattices for superionic water have long been debated.
Initial computational studies~\cite{demontis1988} proposed a face-centered-cubic (fcc) oxygen lattice while early 
first-principles electronic structure molecular dynamics (FPMD) considered superionic water with a body-centered cubic (bcc) oxygen lattice as a high temperature analog of the ice X~\cite{Cavazzoni1999}.
Later FPMD studies re-proposed a fcc~\cite{Wilson2013,French2016}, suggested a close-packed (cp)~\cite{Sun2015}, and, at pressures higher than 1 TPa, a P21/c~\cite{Sun2015} oxygen lattice. 
In the experimental studies of superionic water~\cite{Goncharov2005,Sugimura2012,Millot2018,Millot2019, Queyroux2020},
sample preparation
is extremely challenging,
hydrogen positions cannot be determined,
and temperature measurements in dynamical compression experiments are not straightforward~\cite{Millot2018,Millot2019}. 
Notably, recent dynamical compression experiments combined with x-ray diffraction (XRD) found
a superionic phase with fcc oxygen lattice, ice XVIII, above 2000 K~\cite{Millot2019}.
Static compression experiments combined with synchrotron XRD suggest a triple point between liquid, ice VII$'$, and ice VII$''$ (a bcc superionic phase) at 14.6~GPa and 850~K~\cite{Queyroux2020}.

\begin{figure*}
    \centering
\includegraphics[width=0.99\textwidth]{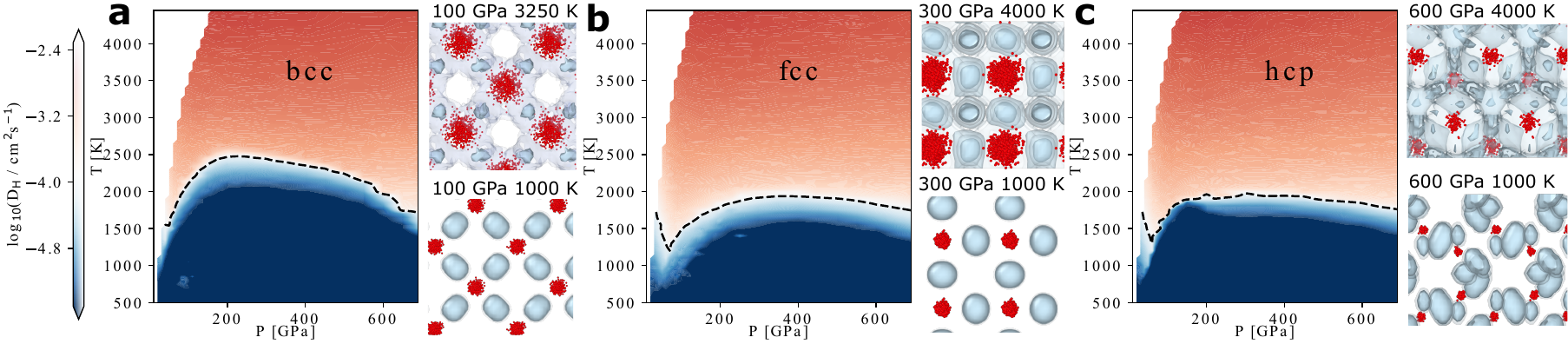}
    \caption{\textbf{The diffusion of hydrogen in water with bcc, fcc, and hcp lattice of oxygen.}\\
    On the left of each panel, the diffusion coefficients of hydrogen $D_\mathrm{H}$ are shown logarithmically as a function of pressure ($P$) and temperature ($T$).
    The dashed lines illustrate the ice to superionic transition temperatures $T_\mathrm{s}$ as defined in the main text.
    On the right, the distributions of hydrogen atoms are indicated using gray contours, and the instantaneous positions of oxygen atoms are indicated using red spheres. 
    For bcc, fcc, and hcp lattices, both a low-temperature ice structure (lower) and a high-temperature superionic structure (upper) are shown.
    }
    \label{fig:superionic}
\end{figure*}

Theoretical computation of the high-P water phase boundaries is also difficult. 
No reliable empirical force fields are available for this system.
FPMD simulations are computationally expensive, and are thus confined to short trajectories and small system sizes, 
which may introduce artifacts leading to  
contradictory results on the locations of phase boundaries~\cite{Cavazzoni1999,Schwegler2008,Redmer2011,Wilson2013,Sun2015,French2016,Hernandez2016,Hernandez2018} and diffusivity~\cite{Wilson2013,French2016} between various studies based on the same assumption of the underlying electronic structure. 
Another shortcoming of most FPMD simulations is that the nuclei are treated as classical point masses
even though they are wave-like quantum mechanical particles. 
Yet, for light elements such as hydrogen,
nuclear quantum effects (NQEs) are particularly acute.
Molecular dynamics (MD) combined with the Feynman path integral (PIMD) can be used to treat the NQEs, but PIMD 
multiplies the computational cost by about another 20-fold~\cite{ceriotti2016nuclear}. 
However, machine-learning potentials (MLPs) can help overcome these limits
by first learning a data-driven model of atomic interactions from first-principles calculations~\cite{Deringer2019}, and then driving large-scale simulations at an affordable computational cost. 
Thanks to their low cost and high accuracy, MLPs have helped reproduce the low-pressure phase diagram of water~\cite{Cheng2019,reinhardt2021quantum},
as well as elucidate the nucleation behaviour of gallium~\cite{Niu2020}, the liquid-liquid transition of high-pressure hydrogen~\cite{cheng2020evidence}, and the structural transition mechanisms in disordered silicon~\cite{deringer2021origins}.

In this study, we 
construct a MLP for high-pressure water using an artificial neural network architecture~\cite{behler2007generalized} based on Perdew-Burke-Ernzerhof (PBE)~\cite{Perdew1996} density functional theory (DFT).
Combining the MLP with advanced free energy methods,
we predict the properties of superionic and liquid water at the PBE level of theory,
using large system sizes, long time scales and considering NQEs.
We elucidate the mechanisms for ice-superionic transitions and hydrogen transport,
map the high-pressure water phase diagram,
and probe the kinetics of phase transition.

\subsection{Hydrogen diffusivity in superionic phases}

To probe hydrogen diffusion, the aforementioned bcc and fcc lattices of oxygen,
and a hexagonal close-packed (hcp) lattice that has a low-T ice analog with the Pbcm space group (structures are in the \si) are considered.
The insulating structures become superionic when the temperature rises,
as marked by the fast diffusion of hydrogen atoms inside the oxygen lattices.
The distributions of position of hydrogen are shown in the contour plots of Fig.~\ref{fig:superionic}:
at low $T$ hydrogen atoms are confined to their equilibrium sites,
while at high $T$ only a fraction occupy these sites at any given time.

The diffusivity of hydrogen $D_\mathrm{H}$ is shown for a wide range of thermodynamic conditions in Fig.~\ref{fig:superionic}.
At about 2000~K, there is a rapid increase of $D_\mathrm{H}$, 
which marks the insulating ice-superionic transition. The dashed lines in Fig.~\ref{fig:superionic} indicate the associated transition temperatures $T_\mathrm{s}$ defined by a cutoff of $10^{-4}$~cm$^2/$s in $D_\mathrm{H}$.
The bcc phase has a higher $T_\mathrm{s}$ compared to fcc and hcp.
At $T\gg T_\mathrm{s}$,
the H diffusion coefficients in all three superionic phases show no distinct difference,
which is in agreement with Ref.~\cite{French2016} and in contrast with Ref.~\cite{Wilson2013}.

We analyze the hydrogen diffusion in the fcc lattice, as the results for bcc and hcp are similar (\si 
).
At pressure and temperature conditions where the lattice remains stable, %
the diffusivity of hydrogen changes rapidly but smoothly across the ice-superionic transition region, as evident in Fig.~\ref{fig:D}a that shows $D_\mathrm{H}$ (on the log scale in the inset) with respect to $T$.
Above $T_\mathrm{s}$, when the system is fully superionic,
$D_\mathrm{H}$ increases gradually as a function of temperature, and is only weakly dependent on pressure (see the \si).

The detailed and microscopic picture of the hydrogen motion revealed in Fig.~\ref{fig:superionic} complements experiments, which cannot determine the positions of hydrogen at the high $P,T$ conditions.
The H diffusion coefficients can be used to derive 
the ionic conductivity of superionic water using a generalized Einstein formula~\cite{French2010},
which is a key quantity in magnetic dynamo simulations for reproducing the complex magnetic field geometry of the Uranus and Neptune~\cite{Soderlund2020}.

\subsection{A physical model of ice-superionic transitions}

To rationalize the ice-superionic transition and the hydrogen diffusion, we start from an interstitial formation model~\cite{rice1974superionic}:
Taking $x$ to be the fraction of the conducting hydrogen atoms,
the free energy of the system can be expressed as
\begin{multline}
f(x,T) = (\epsilon_0- T s_0)x - \dfrac{\lambda}{2} x^2 \\
    + \kb T \left[x\ln x+(1-x)\ln(1-x)\right],
    \label{eq:fx}   
\end{multline}
where $s_0$ is the entropy gain for creating a free interstitial atom in the unit cell from a confined atom, 
as the accessible volume of the former is higher,
and $\epsilon_0$ and $\lambda$ are the energy scales for interstitial formation and interaction, respectively.
Upon equilibration, Eqn.~\eqref{eq:fx} reaches a minimum and
the equilibrium fraction $x(T)= \argmin_x f(x,T)$.

\begin{figure}
    \centering
\includegraphics[width=0.4\textwidth]{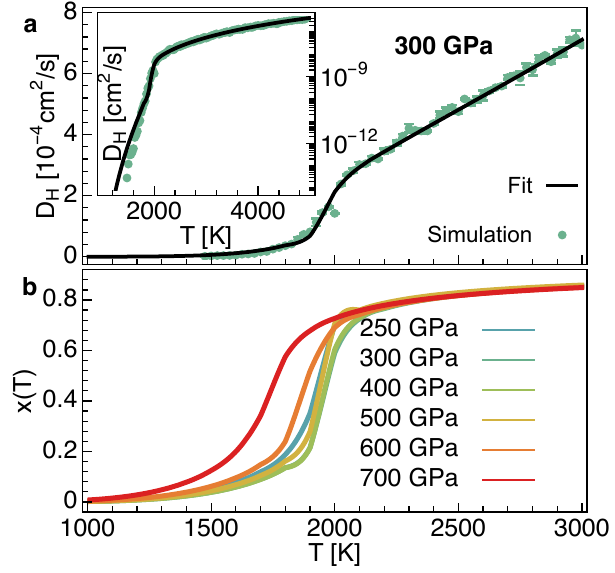}
    \caption{\textbf{Modelling the diffusion behaviors of hydrogen in water phases with a fcc lattice of oxygen.}\\
    Panel \textbf{a}: The diffusion coefficients of hydrogen $D_\mathrm{H}$ at 300~GPa. The black curves are the fits to $x(T)D$.\\
    Panel \textbf{b}: The equilibrium fraction of conducting hydrogen atoms $x(T)$ at different pressures. 
    }
    \label{fig:D}
\end{figure}

To model the diffusivity of conducting hydrogen atoms in the lattice,
we use a Speedy-Angell power-law~\cite{sacco2000temperature}:
\begin{equation}
    D = D_0 \left[ (T/T_0)-1\right]^{\nu},
    \label{eq:d-SA}
\end{equation}
which describes well both the H diffusivity data for the high-pressure water at $T\gg T_\mathrm{s}$ in our simulations and 
water at ambient pressure~\cite{sacco2000temperature}.
Below $T_\mathrm{s}$, we assume that the diffusivity of the conducting hydrogen atoms follows Eqn.~\eqref{eq:d-SA} while the rest have negligible diffusivity, so that overall $D_\mathrm{H}(T) = x(T)D$.
We fit this to the diffusion coefficients computed from the simulations using the MLP,
and one example is shown as the black curve in Fig.~\ref{fig:D}a (all fits are in the \si).
Fig.~\ref{fig:D}b shows $x(T)$ from the fits at different pressures,
which exhibits rapid increase at T$\approx$2000~K corresponding to
the ice-superionic transitions.
At $T > T_\mathrm{s}$, $x(T)$ increases slowly and reaches about $0.8$ at $3000$~K,
consistent with the observation in Fig.~\ref{fig:superionic} that a finite fraction of hydrogen atoms are close to the equilibrium sites even in fully superionic phases.
The interstitial model (Eqn.~\eqref{eq:fx}) combined with the diffusivity expression (Eqn.~\eqref{eq:d-SA})
describe extremely well the simulation results of $D_\mathrm{H}$ over the entire range of thousands of kelvin considered.
Eqn.~\eqref{eq:fx} further sheds light on the driving force of the superionic transition:
The entropy gain $s_0$ competes with the energetic cost $\epsilon_0$ of interstitial formation and wins at high $T$.
The interaction between interstitials $\lambda$ affects the nature of the transition:
$x(T)$ will exhibit a smooth crossover, which is the case observed for the superionic phase here, when $\lambda \le 4\epsilon_0/(2+s_0)$,
and a first-order phase transition in $T$ otherwise.

\subsection{Chemical potentials of superionic and liquid water}

\begin{figure}
    \centering
\includegraphics[width=0.4\textwidth]{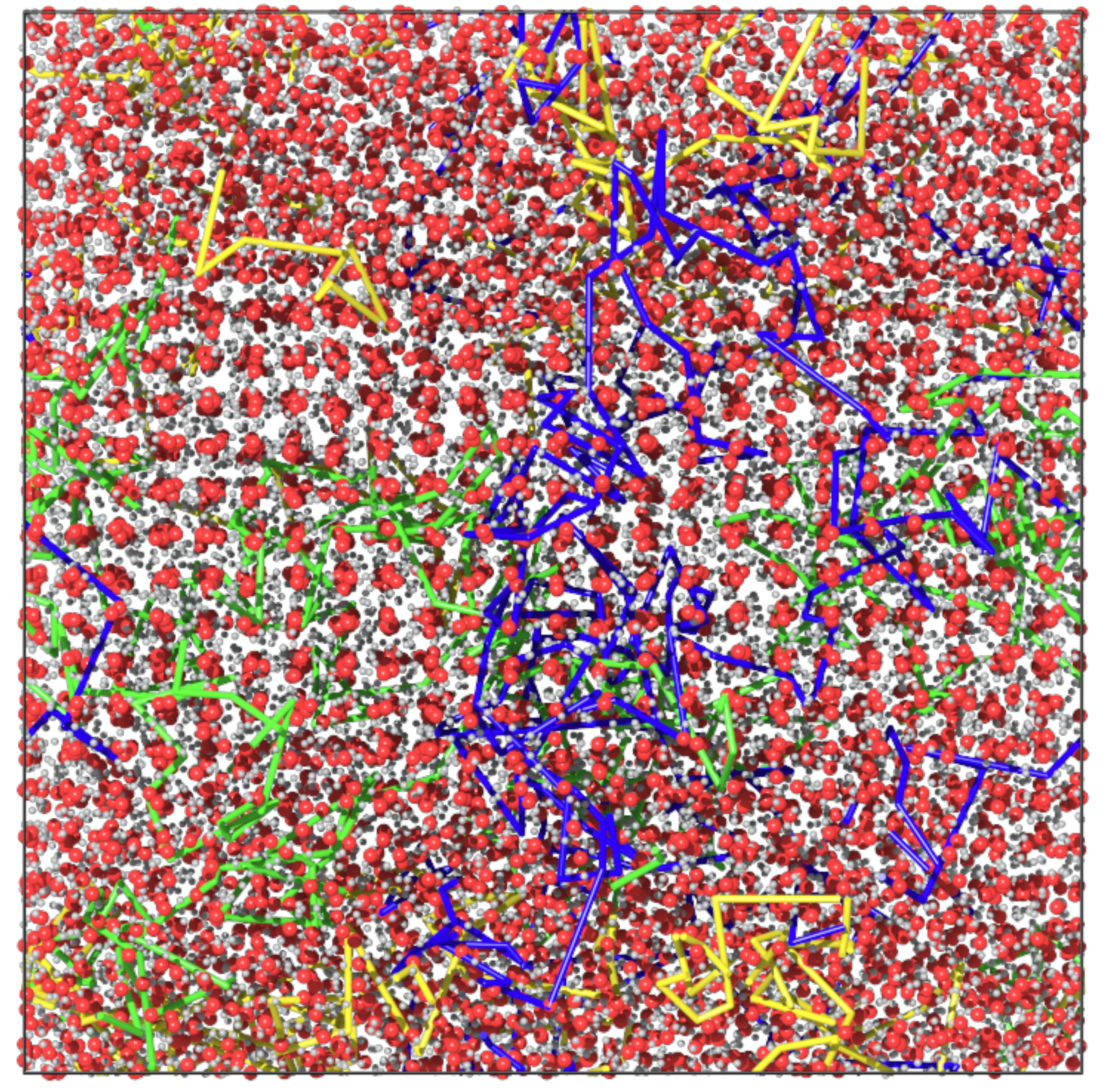}
    \caption{\textbf{Hydrogen atoms diffuse easily between the superionic and the liquid phases. }
    Liquid-superionic(fcc) interface of a water system with 20,736 atoms at 100~GPa and 3250~K (on the melting line). The oxygen atoms are in red and the hydrogen atoms are in white. 
    The yellow, green and blue lines show the trajectories of three hydrogen atoms during a 75~ps molecular dynamic simulation run using the MLP.
    }
    \label{fig:sl}
\end{figure}

\begin{figure*}
    \centering
    \includegraphics[width=0.95\textwidth]{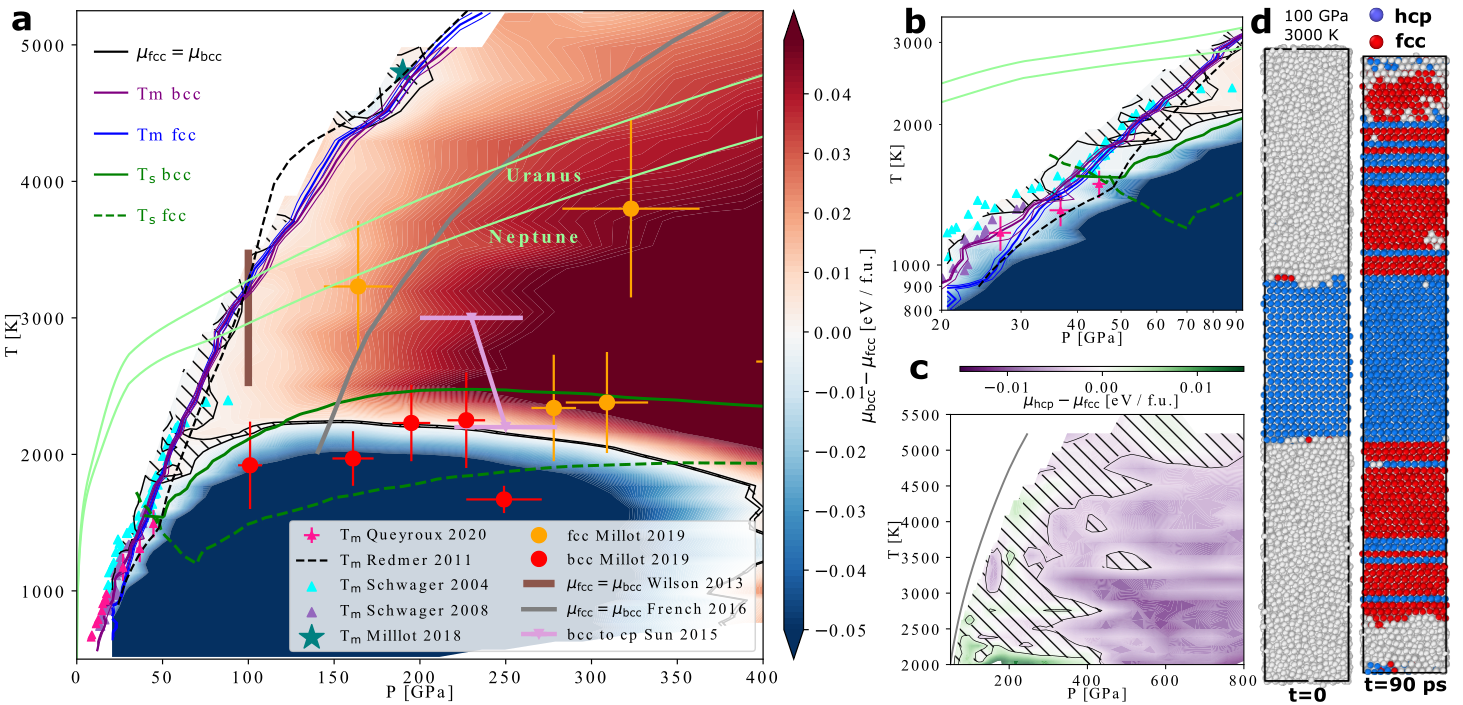}
    \caption{
    \textbf{The phase stabilities for bcc, fcc, hcp and liquid water.}\\
    Panel \textbf{a}: The chemical potential difference $\mu_\mathrm{bcc}-\mu_\mathrm{fcc}$ per formula unit (f.u.) between bcc and fcc at different pressure ($P$) and temperature ($T$) conditions.
    The blue and the purple lines are the melting lines $\tm$ for fcc and bcc, respectively, with the statistical uncertainties indicated by the upper and lower thin lines.
    The green lines are the superionic-ice transition lines $T_\mathrm{s}$ 
    for fcc (dashed lines) and bcc (solid lines).
    The hatched area indicates the estimated region of $\mu_\mathrm{bcc}=\mu_\mathrm{fcc}$ with statistical uncertainty.
    Experimental results are from Schwager~2004~\cite{Schwager2004}, Schwager~2008~\cite{Schwager2008},
    Millot~2018~\cite{Millot2018},
    Millot~2019~\cite{Millot2019}, 
    and Queyroux~2020~\cite{Queyroux2020},
    and theoretical predictions are from
    Redmer 2011~\cite{Redmer2011},
    Wilson~2013~\cite{Wilson2013}, 
    Sun~2015~\cite{Sun2015}, and French~2016~\cite{French2016}.
    Planetary interior conditions for
    Neptune and Uranus (bright green lines) are from Ref.~\cite{Scheibe2019}.\\
    Panel \textbf{b}: A zoom-in at $P<100$~GPa of the panel \textbf{a} with $P$ and $T$ both on the log scale. Legends are shared.\\
    Panel \textbf{c}: The chemical potential difference $\mu_\mathrm{hcp}-\mu_\mathrm{fcc}$ between the hcp and fcc superionic phases. 
    The statistical uncertainties in the coexistence line are indicated by the hatched area.\\
    Panel \textbf{d}: Oxygen lattice growth starting from a hcp slab with closed-packed planes parallel to the interface with liquid water.
    }
    \label{fig:phase-diagram}
\end{figure*}

The thermodynamic stability of different phases is governed by their chemical potentials $\mu$.
However,
computing the $\mu$ of the superionic phases is difficult, 
because these phases are half-solid-half-liquid,
and because the thermodynamic integration (TI) method cannot be applied across first-order boundaries that might exist between a number of phases all sharing the bcc oxygen lattice~\cite{Hernandez2018,Queyroux2020} (e.g. ice VII, VII$'$, VII$''$, X).
To circumvent these difficulties, we instead compute their relative chemical potentials to the liquid, $\mu^\mathrm{MLP}_\mathrm{fcc/bcc/hcp}-\mu^\mathrm{MLP}_\mathrm{l}$,
in
umbrella sampling~\cite{torrie1977nonphysical} simulations on superionic-liquid coexistence systems (illustrated in Fig.~\ref{fig:sl}) at different pressures between 15 and 200~GPa and temperatures close to melting points using the MLP.
In these coexistence systems, oxygen atoms in the liquid phase diffuse around, and the ones in the superionic phase stay on the bcc, fcc or hcp lattices.
Meanwhile, hydrogen atoms travel in and out between the two phases,
as illustrated by a few %
H trajectories in Fig.~\ref{fig:sl}.

From the chemical potentials relative to the liquid phase at a given $(P,T)$, the difference in $\mu$ between two superionic phases is just, e.g. $\mu^\mathrm{MLP}_\mathrm{bcc}-\mu^\mathrm{MLP}_\mathrm{fcc} =
(\mu^\mathrm{MLP}_\mathrm{bcc}-\mu^\mathrm{MLP}_\mathrm{l})-(\mu^\mathrm{MLP}_\mathrm{fcc}-\mu^\mathrm{MLP}_\mathrm{l})$. 
We then use TI along isotherms and isobars to calculate the chemical potential differences at any other conditions where the two phases remain metastable.
After that,
we promote the MLP results to the PBE DFT level by adding $\mu-\mu^\mathrm{MLP}$ computed from free-energy perturbation,
which removes the small residual errors in the MLP partly due to its lack of long-range electrostatics~\cite{behler2015constructing,Cheng2019}.
Finally, we include NQEs by performing PIMD simulations using the MLP, and finally obtain $\mu_\mathrm{fcc/bcc/hcp}$ at the PBE DFT level that fully take into account quantum thermal fluctuations.

The melting curves ($\tm$) of bcc and fcc are indicated using the purple and the blue lines in Fig.~\ref{fig:phase-diagram}a,b, respectively.
They are similar to each other at $P\ge40$~GPa, 
and noticeably different from previous single-phase melting and solidification FPMD calculations~\cite{Redmer2011}.
At $P<100$~GPa (Fig.~\ref{fig:phase-diagram}b), our $\tm$
are roughly in between 
the experiments based on laser-heated DAC with XRD at $P\le45$~GPa~\cite{Queyroux2020}, and the direct observation 
of motion in the laser-speckle pattern
at $P\le$ 90 GPa~\cite{Schwager2004,Schwager2008}.
Our $\tm$ are presented up to 220 GPa, and agree with a single point at about 190~GPa and 4800~K in the precompressed shockwave experiment~\cite{Millot2018}.

The chemical potential difference between bcc and fcc is plotted in Fig.~\ref{fig:phase-diagram}a. 
The bcc lattice is stable at low $T$ as well as low $P \lesssim$ 75~GPa.
The two black curves show the lower and the upper bounds of the coexistence temperature $T(\mu_\mathrm{fcc}=\mu_\mathrm{bcc})$,
and the hatched area indicates the statistical uncertainty of the coexistence line.
The region of stability of the bcc lattice includes both the insulating ice and the superionic bcc, which are separated by the aforementioned ice-superionic transition line $T_\mathrm{s}$ (the solid green line in Fig.~\ref{fig:phase-diagram}a).
The superionic bcc phase has a narrow stability region,
and becomes progressively less stable at higher $P$.
Previous calculations of the boundary between bcc and close-packed (cp) superionic phases were conflicting and scattered:
TI calculations using FPMD at 3000~K~\cite{Wilson2013} (the brown line),
constant-pressure single-phase FPMD simulations at 2200~K and 3000~K~\cite{Sun2015} (the pink line with the upper and lower bounds of transition pressure between bcc and cp phases),
and analytic models combined with FPMD~\cite{French2016} (the gray curve).
Our boundary has small statistical uncertainty thanks to the large-scale simulations. 
Notably, our boundary is fully compatible with the pressure–temperature conditions where bcc (red circles) and fcc (orange circles) phases were observed in
the XRD measurements~\cite{Millot2019}.

The chemical potential difference between hcp and fcc is plotted in Fig.~\ref{fig:phase-diagram}c.
hcp becomes more favorable at higher pressure, and is slightly more stable than fcc at above $\approx$400~GPa.
Overall, the magnitude of $\mu_\mathrm{hcp}-\mu_\mathrm{fcc}$ is very small across 100~GPa $\le P \le$ 800~GPa and 2000~K$\le T \le$~5500~K, of less than 10~meV per H$_2$O formula unit, compared with the thermal energy of $0.3$~eV per degree of freedom at a few thousands of kelvin.
This hints that fcc and hcp can probably coexist at these conditions, which has been observed in FPMD simulations at $P>280$~GPa in Ref.~\cite{Sun2015}.
To further confirm this, we simulate superionic water growing from liquid water supercooled at $3000$~K, $100$~GPa (shown in Fig.~\ref{fig:phase-diagram}d).
The starting configuration has a pure hcp oxygen lattice with close-packed planes
parallel to the superionic-liquid interface,
and the oxygen lattice gradually grows into a state of mixed stacking.
This provides strong indication that the equilibrium phase at the fcc region of stability (shown in Fig.~\ref{fig:phase-diagram}a) may have a finite fraction of mixed stacking of close-packed planes.
Such stacking can be revealed from XRD measurements,
and we provide the simulated diffraction patterns (in the \si) as a guide.

\subsection{Superionic-liquid interfacial free energies}

Characterising kinetic effects in phase transition is pivotal in interpreting experiments,
because
the phase synthesized may not be the stable phase suggested by the phase diagram (Fig.~\ref{fig:phase-diagram}a,c), 
but a metastable phase with lower activation barrier to nucleate~\cite{VanSanten1984}.
The interfacial free energy $\gamma$ has a dominant influence on nucleation,
as it enters the nucleation rate through an exponential of a cubic power~\cite{oxtoby1992homogeneous}.
$\gamma$ is also a key parameter in hydrodynamic simulations of dynamical compression experiments~\cite{Myint2018,Myint2020}.
We compute fcc/bcc-liquid interfacial free energies $\gamma$ for the $\avg{100}$, $\avg{111}$ and $\avg{110}$ interfaces (see Table~\ref{tab:gmtm}) at $T=3250$~K and $P=100$~GPa (on the $\tm$) using the capillary fluctuation method (see \si~and Ref.~\cite{davidchack2006anisotropic}).
We assume that hcp is likely to have similar $\gamma$ as fcc, due to their shared close-packing.
As both fcc and bcc observe cubic symmetry,
the orientation-dependent $\gamma(\vec{n})$ can be expanded using a cubic harmonic series~\cite{davidchack2006anisotropic}:
\begin{multline}
        \gamma(\vec{n})/\gamma_0 =
    1 + \epsilon_1 \left(\Sigma_{i=1}^3 n_i^4 -\dfrac{3}{5}\right)\\
    + \epsilon_2 \left(3\Sigma_{i=1}^3 n_i^4 + 66n_1^2n_2^2n_3^2-\dfrac{17}{7}\right),
\end{multline}
where $\{n_1 ,n_2 ,n_3\}$ are the Cartesian components of the unit
vector $\vec{n}$ normal to the interface, $\gamma_0$ is the orientationally
averaged interfacial energy, $\epsilon_1$ and $\epsilon_2$ quantify the anisotropy (see Table~\ref{tab:gmtm}). 
Both bcc and fcc have small anisotropy in $\gamma(\vec{n})$, suggesting that the equilibrium shapes of superionic nuclei in liquid water are near-spherical.

The bcc superionic phase has much lower $\gamma_0$ ($16.8 \pm 0.2~$meV$/\AA^2$) compared to fcc ($24.4 \pm 0.3$~meV$/\AA^2$) at the given condition,
suggesting that the former is easier to nucleate.
This means that, when the kinetics of phase transitions play a significant role,
the bcc superionic phase may form even at conditions where it is not as stable as fcc.

Crucially, planets have billions of years to evolve and to reach equilibrium, but dynamic compression experiments~\cite{Millot2018,Millot2019},
pulsed laser heating in the DAC~\cite{Goncharov2008,prakapenka2020polymorphism}
and FPMD simulations have short time scales.
The kinetic factors probed here help bridge the gap between the thermodynamics of water inside the planets and the observations from these fast experiments and simulations.

\begin{table}
\caption{A comparison of the computed interfacial free energy and anisotropy at $\tm$ for bcc and fcc superionic phases.
The numbers in brackets indicate the statistical uncertainties of the last digit.}
\label{tab:gmtm}
    \begin{tabular}{ c | c c c | c c c}
   \hline\hline
  & \multicolumn{3}{ c| }{$\gamma$[meV$/\AA^2$]} & \multicolumn{3}{ c }{parameters} \\
  & $\gamma_{100}$ & $\gamma_{111}$ & $\gamma_{110}$
  & $\gamma_0$[meV$/\AA^2$] & $\epsilon_1$ & $\epsilon_2$ \\
  
  \hline   
  fcc & 
  24.5(4) & 24.1(4) & 24.2(4) &
  24.4(3) & 0.0252(4) & 0.0002(8)
  \\
  bcc &
  16.7(3) & 17.0(2) & 16.8(2) &
  16.8(2) & -0.0198(4) & 0.003(1)
 \\ \hline \hline
    \end{tabular}
\end{table}

\section{Conclusions}

To conclude, we predict the dynamics, thermodynamics and kinetics of superionic water with unprecedented range and resolution.
We do so by running large-scale molecular dynamics simulations using a machine learning potential, combined with PIMD and free-energy techniques.
Compared with previous experimental and theoretical data shown in Fig.~\ref{fig:phase-diagram}, which are patchy and contradictory,
we are able to quantitatively map the behaviors of superionic water cross a large part of the phase diagram (10~GPa~$\le P \le$~800~GPa, 500~K~$\le T \le$~5500~K), along with mechanistic understanding of the ice-superionic transition.

Our predictions of the phase boundaries of water at planetary conditions can be used to determine the fraction of insulating ice, superionic, and liquid water in the interior of ice giants.
Our results suggest that a close-packed superionic phase with mixed stacking is stable over a wide temperature and pressure range (Fig.~\ref{fig:phase-diagram}),
while the superionic bcc phase is stable in a small region at $P\lesssim 75$~GPa and $T\lesssim 2500$~K, but may be kinetically favored due to its lower interfacial free energy with the liquid.
The planetary interior models illustrated in Fig.~\ref{fig:phase-diagram}a combined with our phase diagram imply a transition from liquid water to superionic water deep inside the ice giants.
This transition zone will see a liquid-superionic water interface, with hydrogen atoms diffusing across the interface between the two phases like what we see in Fig.~\ref{fig:sl}.
These trespassing hydrogen atoms will conduct electrical charges and heat across such interfaces in the icy planets.

Our quantitative understanding of superionic water sheds light into the interior structure, evolution and the dynamo process of planets such as Uranus and Neptune and also of the increasing number of icy exoplanets~\cite{Zeng2019}.
We suggest future static and dynamic compression experiments to discover the close-packed phase with mixed stacking, and to investigate the preferential nucleation of bcc.
Furthermore, our framework can be used to discover and characterize 
superionic electrolyte materials,
as well as new superionic phases of other components such as methane, ammonia, salts and related mixtures that are relevant for planetary science.

\textbf{Acknowledgements}
We thank Marius Millot, Aleks Reinhardt, Guglielmo Mazzola for reading an early draft and providing constructive and useful comments and suggestions. We also gratefully acknowledge Jean-Alexis Hernandez and Ludwig Scheibe for sharing their equation of state data and planetary profiles.
BC acknowledges resources provided by the Cambridge Tier-2 system operated by the University of Cambridge Research Computing Service funded by EPSRC Tier-2 capital grant EP/P020259/1. CJP acknowledges support from the EPSRC grant EP/P022596/1. MB was supported by the European Horizon 2020 program within the Marie Sk{\l}odowska-Curie actions (xICE grant number 894725).
Part of this work was performed under the auspices of the U.S. Department of Energy by Lawrence Livermore National Laboratory under Contract DE-AC52-07NA27344. LDRD 19-ERD-031 and 21-ERD-005 provided partial support. Computing support for this work came in part from the Lawrence Livermore National Laboratory (LLNL) Institutional Computing Grand Challenge program.

\textbf{Data availability statement}
The authors declare that the data supporting the findings of this study are available within the paper,
and detailed description of the calculations together with all necessary input files to reproduce the reported results are included in Supplemental Information.
All original data generated for the study,
and the machine learning potential for high-pressure water constructed in this study are in the \si~and repository \url{https://github.com/BingqingCheng/superionic-water} (will be made public upon acceptance of the manuscript).

\end{document}